\documentclass{aa}
\usepackage{graphicx}
\usepackage{natbib}
\usepackage{color}

\begin{document}

\title{The first study of the light-travel time effect \\ in massive LMC eclipsing binaries\thanks{Based
  on data collected with the Danish 1.54-m telescope at the ESO La Silla Observatory.}$^,$\thanks{Table 4 is
  only available in electronic form at the CDS via anonymous ftp to cdsarc.u-strasbg.fr (130.79.128.5)
  or via http://cdsweb.u-strasbg.fr/cgi-bin/qcat?J/A+A/}}

\author{P. Zasche
    \and M. Wolf
    \and J. Vra\v{s}til
    \and L. Pilar\v{c}\'{\i}k
    \and J. Jury\v{s}ek
    }

\offprints{Petr Zasche, \email{zasche@sirrah.troja.mff.cuni.cz}}

 \institute{Astronomical Institute, Charles University in Prague, Faculty of Mathematics and
Physics, CZ-180~00 Praha 8, V~Hole\v{s}ovi\v{c}k\'ach 2, Czech Republic }

\date{Received \today}

\titlerunning{The first study of LTTE in LMC eclipsing binaries.}
\authorrunning{P. Zasche et al.}

\abstract{}
 {New CCD observations for semidetached and detached eclipsing binaries from the Large Magellanic
Cloud were carried out using the Danish 1.54-meter telescope located at the La Silla Observatory in
Chile. The selected systems were monitored for their times of minima, which were required to be able to
study the period changes taking place in them. In addition, many new times of minima were derived from the
photometric surveys OGLE-II, OGLE-III, and MACHO.}
 {The $O-C$ diagrams of minima timings were analysed using the hypothesis of the light-travel time effect,
i.e. assuming the orbital motion around a common barycenter with the distant component. Moreover,
the light curves of these systems were also analysed using the program {\sc PHOEBE}, which provided the
physical parameters of the stars.}
 {For the first time, in this study we derived the relatively short periods of modulation in these systems, which
relates to third bodies. The orbital periods resulted from 3.6 to 11.3~years and the eccentricities
were found to be up to 0.64. This is the first time that this kind of analysis for the set of
extragalactic sources has been performed. The Wolf-Rayet system OGLE-LMC-ECL-08823 is the most
mysterious one, owing to the resultant high mass function. Another system, OGLE-LMC-ECL-19996, was
found to contain a third body with a very high mass ($M_{3,min}=26$~M$_\odot$). One system
(OGLE-LMC-ECL-09971) is suspicious because of its eccentricity, and another one
(OGLE-LMC-ECL-20162) shows some light curve variability, with a possible flare-like or
microlensing-like event.}
 {All of these results came only from the photometric observations of the systems and can be
considered as a good starting point for future dedicated observations.}

\maketitle

\section{Introduction}

Monitoring the eclipsing binaries and their eclipses is a classical technique in stellar
astrophysics, which has been used for decades. One can study the eclipsing binary light curve and
model its shape, which reveals the physical parameters of both eclipsing components, as well as its
orbit (see e.g. \citealt{2009ebs..book.....K}). It is still the most precise method to derive the
masses, radii, and luminosities of components (see e.g. \citealt{2012ocpd.conf...51S}).

The role of the eclipsing binaries outside of our Galaxy is undisputable, however their
observations have been quite problematic owing to their low brightness and therefore a lack of data
for analysis. This issue has changed rapidly during the last two decades thanks to the large
surveys like OGLE and MACHO. Owing to the long-lasting photometric monitoring of the Magellanic
Clouds, nowadays we
 know more than 30 000 eclipsing binaries outside of the Galaxy, and many of them are
interesting enough for further, more detailed analysis. Hence our contribution to the topic is
still valuable.

The study of eclipsing binaries as extragalactic sources was also motivated for another reason. The
Magellanic Clouds have slightly different metallicity than our Milky Way Galaxy (see e.g.
\citealt{1997macl.book.....W}, or \citealt{2015ApJ...806...21D}), hence one can study whether this
effect plays a role in eclipsing binary research, whether it is traceable in the models, or
whether our data are sufficiently precise. Another motivation can also be the question of the
stellar multiplicity in general. Is the frequency of multiples about the same in Magellanic Clouds
as in our own Galaxy? \cite{2016MNRAS.455.4136B} show quite recently that of about 1/12 of all
eclipsing binaries observed by the Kepler space telescope probably contain additional components
that can be detected only via eclipse timing variations.

\section{The system selection}

Over the last few years, we performed an analysis of several interesting binaries located in the
Magellanic Clouds, which show the apsidal motion (see e.g. \citealt{2015AJ....150..183Z}). All of
these binaries are well-detached, slightly eccentric and have orbital periods of the order of
couple of days. Thanks to this long-term monitoring of the selected fields, we also collected many
useful photometric observations for the stars close to our main targets and checked which of them
are adequately interesting for future analysis.

The stars were also checked for their light curves in the {\sc Macho} \citep{2007AJ....134.1963F},
{\sc OGLE II} \citep{2004AcA....54....1W}, and {\sc OGLE III} \citep{2011AcA....61..103G}
databases. The data mining from all these data sources, together with our new photometry led to the
selection of several quite interesting systems, where the suspicious behaviour of the minima times
was found. This was the first selection criterion.

Another selection criterion was the data coverage. Because we focused on periodic variations in the
$O-C$ diagrams, we decided to include in our study only systems that have at least one period of
the variation already covered (either with the survey data or our own observations). All of the
selected stars are of detached or semidetached type.

\section{The analysis}

For the analysis of the light curves, we used the program {\sc PHOEBE} \citep{2005ApJ...628..426P},
which is based on the Wilson-Devinney algorithm \citep{1971ApJ...166..605W}. We used the OGLE III
data for the light curve modelling, because these are of the best quality, obtained over a long
time span and the phase light curves are well covered.

The {\sc PHOEBE} code enables us to construct the theoretical light curve, which is later used as a
template to derive the times of eclipses. Therefore, our light curve fit needs to be as precise as
possible. For the light curve fitting, we assumed all the systems were circular, hence the
eccentricity was fixed at zero. For the starting ephemerides, we used the same ones as published by
\cite{2011AcA....61..103G} in their catalogue. The primary temperatures were derived from the
photometric indices as given by \cite{2012AcA....62..247U}, \cite{2002ApJS..141...81M}, and
\cite{2004AJ....128.1606Z}. See Table \ref{InfoSystems} for more information about the individual
systems in our sample.

Therefore, the set of the fitted quantities was the following: The temperature of the secondary
component $T_2$, the inclination angle $i$, the Kopal's modified potentials $\Omega_i$, and the
luminosities of the components $L_i$. Having no information about the radial velocities of the
eclipsing pair, the mass ratio value was derived in the following way. For all of the systems, we
tried two different configurations - detached as well as semidetached ones. For the detached one we
used the method of deriving the mass ratio as introduced in \cite{2003MNRAS.342.1334G}, using the
assumption that both components are located on the main sequence. For the semidetached ones, the
mass ratio was directly computed as a free parameter. For most of the systems the hypothesis of the
semidetached configuration led to slightly worse fits with higher values of the $\chi^2$ than the
detached one. The limb darkening coefficients were interpolated from the van Hamme's tables
\citep{1993AJ....106.2096V} and the synchronicity parameters ($F_i$) were also kept fixed at values
of $F_i = 1$. Because we deal with very hot stars here, we also fixed the albedo coefficients $A_i$
at a value 1.0, as well as the gravity darkening coefficients $g_i = 1.0$. And finally, the value
of the third light was also computed during the light curve solution. This luminosity $L_3$ cannot
necessarily be connected with the eclipsing pair itself (i.e. only an optical double), but here we
naturally explain its origin by the third-body hypothesis in the system and directly identify this
contribution with the third hidden component. For the results of the fitting process, see the
parameters given in Table \ref{LCparam}.

To study the period variations in these binaries, we used a well-known light-travel time effect (or
LTTE) hypothesis \citep{Irwin1959}. This is based on the assumption that there are not only the two
stars of the binary, but also a hidden distant third component, orbiting around the eclipsing pair.
As the pair moves around a common barycenter of the system, the eclipses of the binary occur
earlier or later depending on the current position of the binary with respect to the observer. For
a discussion and limitations of the method see e.g. \cite{Mayer1990}. A similar method was recently
used to discover several dozens of new triple systems in the Kepler data set (see
\cite{2016MNRAS.455.4136B} or \cite{2015AJ....150..178G}). Analysis of many eclipsing binaries in
the LMC fields that were observed by our group led to an identification of a few interesting ones,
which show some periodic modulation of their orbital period. These systems are now presented in
this study in more detail.

During the last few decades, it has been found that most of the early-type stars are multiples (see
e.g. \citealt{Chini2012} or \citealt{2013ARA&A..51..269D}). And obviously, all of the stars in our
sample are rather massive stars (spectral type B and earlier), hence the LTTE should be detected
for many of them. Surprisingly, there is still a lack of studies of the period changes and
detection of the third components for the stars in the Magellanic Clouds. Besides a few studies of
the apsidal motion in eccentric eclipsing binaries, only one paper was found about the period
variations for the systems in the Large Magellanic Cloud \citep{2015PKAS...30..211R}. They state
that, after an analysis of 79 EROS systems, nine apsidal motions, eight mass transfers, and 12 LTTE
systems were identified. But only one LTTE system is being presented in their paper. However, this
one system still has a rather questionable solution with the combination of the mass transfer
together with the LTTE hypothesis with the orbital period of about 100 yr (but the data cover only
20 yr).

The times of minima for the analysed systems were derived using the AFP method presented in
\cite{2014A&A...572A..71Z}. This method uses the light curve template, as derived from {\sc
PHOEBE,} and shifts the template in both $x$ and $y$ axes together with the phased light curve in
the particular dataset to achieve the best fit. These datasets were constructed according to the
quality and density of the data (for OGLE this usually means one dataset per one year of
observations). Using the MACHO, OGLE, and our new data we obtained minima times spread out over
20-yr period for some of the systems. Hence, detecting the orbital periods of the order of several
years was possible. All of these data points are given in the appendix tables, see Table
\ref{MINIMA} as an example.

The whole fitting process was performed in several steps. At first the ephemerides from
\cite{2011AcA....61..103G} were used and the preliminary solution was found in {\sc PHOEBE}. With
this light curve template, the AFP method produced some preliminary minima and we were able to see
whether the system was suitable for a further analysis or not. The second step was the period
analysis, refining the orbital period, which was then used in {\sc PHOEBE} for a more detailed
modelling of the light curve. With the final light curve template, the final times of minima were
derived and the period analysis was performed.

One can ask what kind of phenomena can be studied in this way and about the suitability of the
method of $O-C$ diagram analysis for the eclipsing binaries located in the LMC. A classical
explanation of the period changes can be found in various literature published in recent decades,
hence we only provide a summary. The periodic modulations of the orbital period are usually
explained via the third-body hypothesis and the light-travel time effect (i.e. only the geometrical
effect, see above). On the other hand, a dynamical effect of perturbations also exists, which is
caused by the third component(s), see, e.g. \cite{2015MNRAS.448..946B}. However, its contribution
is only large enough for the tight systems where the third body has a sufficiently short orbital
period. The extending, or on the opposite, the shortening of the orbital period of the binary is
typically described as the mass transfer between components (or escaping from the system), see,
e.g. \cite{2001icbs.book.....H}. On the other hand, if we deal with the eccentric orbit and the
primary and secondary minima behave in opposite way, we usually apply the hypothesis of the apsidal
motion of the binary \citep{1983Ap&SS..92..203G}. However, sometimes the changes of period are not
strictly periodic, or are even abrupt. The quasiperiodic modulation is often described via some
magnetic activity of the components (which is preferably present in the later-type stars), see,
e.g. \cite{Applegate1992}, and the discontinuous period changes are usually explained as some kind
of mass ejections from the components \citep{2014ApJ...795....8W}. And, of course, in the real
system, all of these mechanisms can be present together, hence their contributions have to be
summed up. A review on this topic can be found in, for example, \cite{2005ASPC..335....3S}. The
results of our fitting is presented below in Table \ref{OCparam}.

\begin{figure*}
  \centering
  \includegraphics[width=0.9\textwidth]{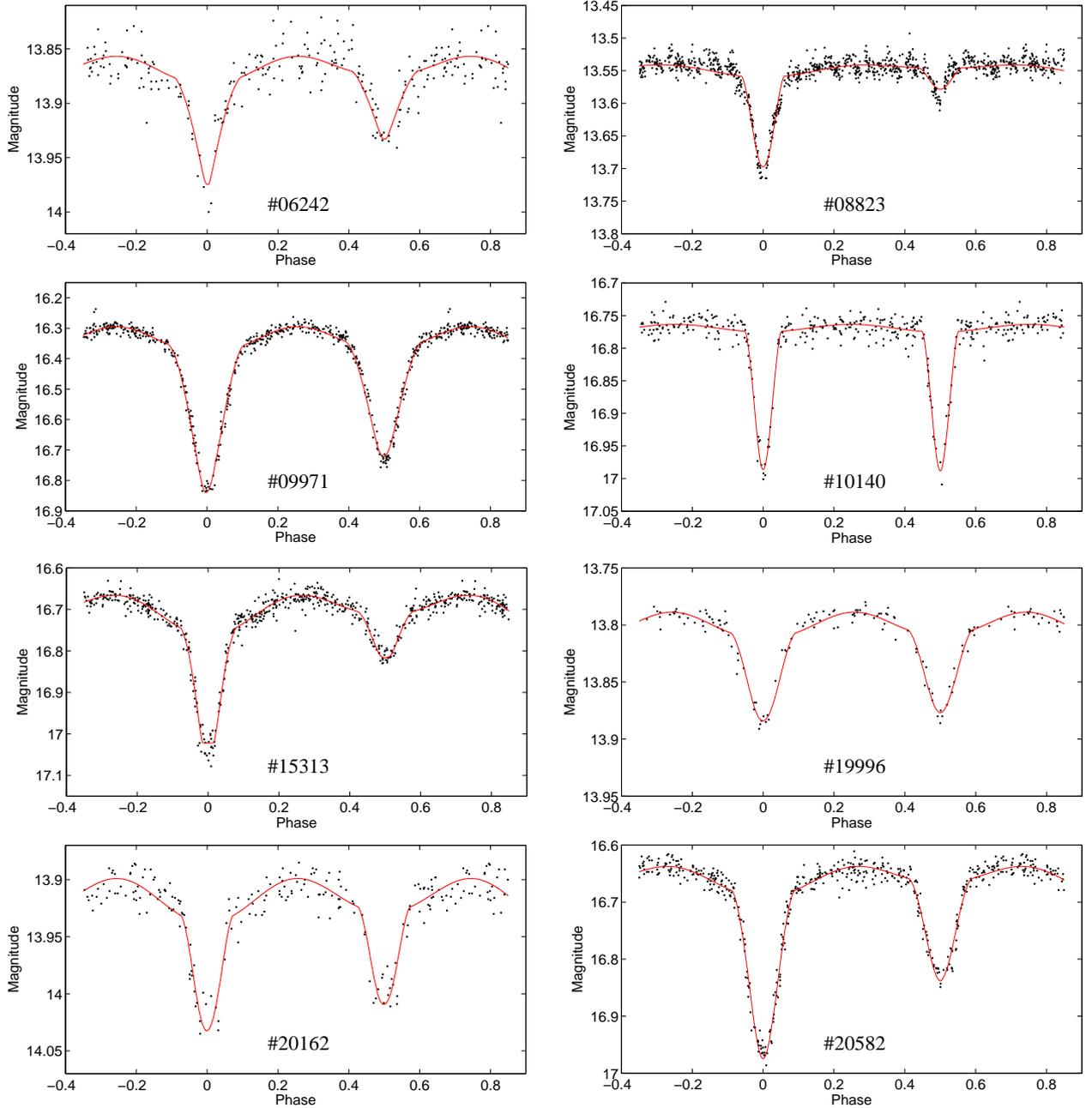}
  \caption{Plot of the light curves of the analysed systems.}
  \label{FigLC}
\end{figure*}

\section{Individual systems}

In the tables and pictures, we decided to use an abbreviation of the long OGLE names. Therefore,
instead of, for example, OGLE-LMC-ECL-06242, we use only the $\#$06242 for increased clarity.

Because the method of analysis is the same for all of the studied systems and the results are
sometimes similar, we focus only on the most interesting ones in our sample and discuss them in
more detail in the following subsections.

\subsection{OGLE-LMC-ECL-08823}

The most interesting system in our sample is definitely OGLE-LMC-ECL-08823, which is also the
hottest one, of earliest spectral type, and also probably the most massive. This is in fact the
only star in our sample that was analysed before. \cite{2003MNRAS.338.1025F} published the study of
the Wolf-Rayet stars in Magellanic Clouds and this star was also included. They find that it is of
WN4b spectral type, while the secondary is probably an O5 star. The effective temperature of 79
000~K, as listed in Table \ref{LCparam}, is taken from this paper. The spectra of the system were
studied, but the secondary radial velocities were not derived, hence we only deal with the SB1-type
binary.

\begin{figure*}
  \centering
  \includegraphics[width=0.9\textwidth]{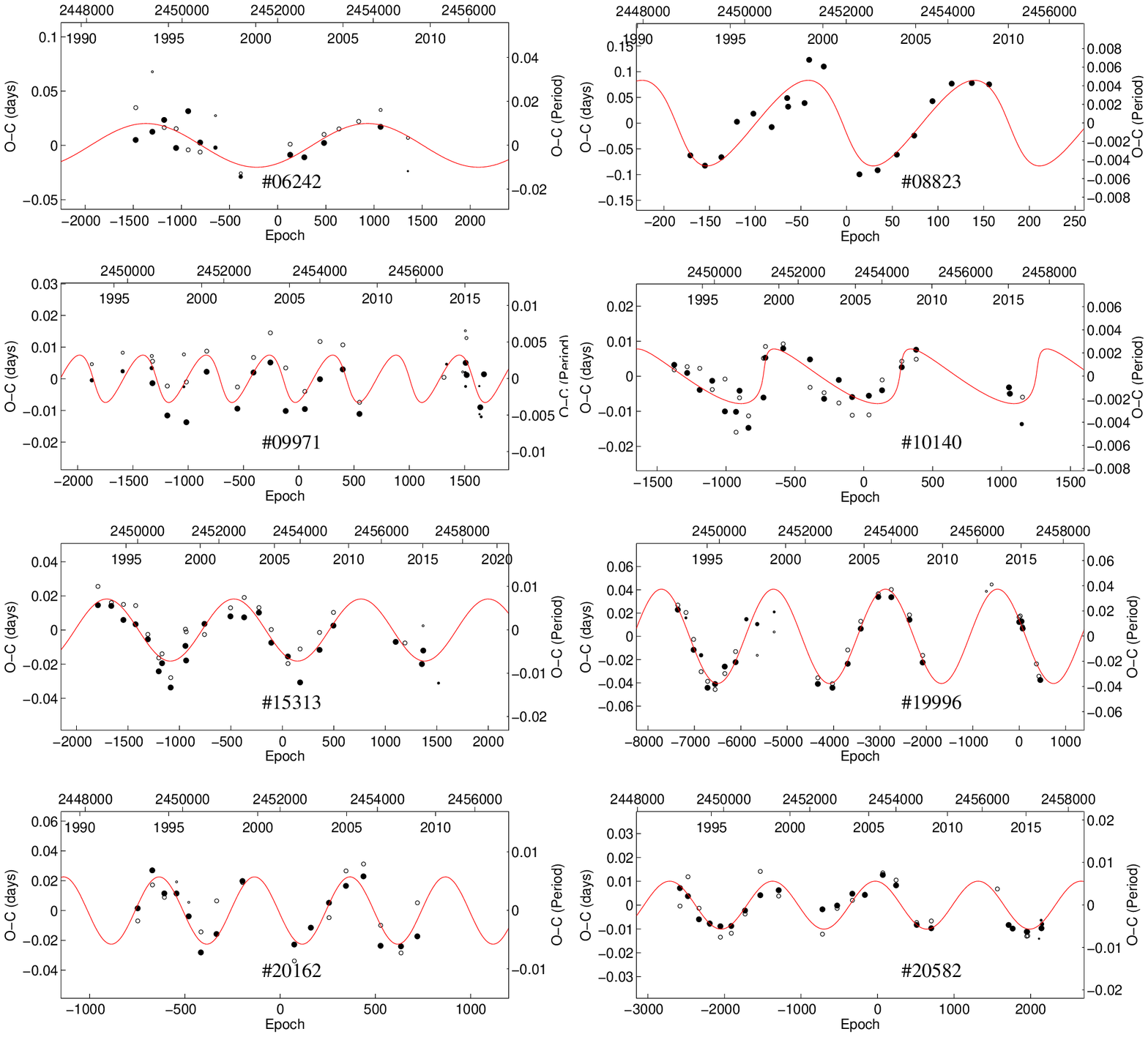}
  \caption{Plot of the $O-C$ diagrams of the analysed systems. The individual primary and
  secondary minima are denoted by dots and open circles, respectively. Larger symbols correspond
  to the more precise measurements. The solid lines indicate the final fit.}
  \label{FigOC}
\end{figure*}

We used OGLE III data for the light curve modelling. As well as this data set, we used OGLE II and
MACHO photometry to determine minima times. With its rather long orbital period of about 18~days,
it is also the longest period star in our study. However, owing to its long period, we were not
able to obtain any new observation of its eclipses, hence the time span of the minima times is only
limited to about 16~yr. For the light curve modelling, we used the assumption that the mass ratio
value is $q=2.67$ (coming from a rough estimation of masses of WN4 and O5V stars). As one can see
in Fig. \ref{FigLC}, the light curve fit is not perfect and the whole light curve is probably
somehow distorted (upper parts of the primary minimum seems to be wider), and the secondary minima
are only shallow. For this reason, we decided to use only the primary minima for the period
analysis. From Fig. \ref{FigOC}, it is clear that the orbital period is modulated with a period of
about 8.9-yr. However, its amplitude is so large, that it produces an unrealistically high value of
the mass function for the third component. Thus, its minimal mass (assuming $M_{12}=55~$M$_\odot$)
also resulted in more than 100~M$_\odot$, which seems to be a rather improbable solution.
Therefore, we have to conclude that, for this system, the LTTE hypothesis probably produces
spurious results, or the period changes are abrupt and caused by another mechanism. However, the
publications on the period changes in Wolf-Rayet stars are still very rare these days.

\subsection{OGLE-LMC-ECL-19996}

The second very interesting case is OGLE-LMC-ECL-19996, which is also one of the stars with the
earliest spectral type in the LMC. This star has not been studied before and only the photometry
from the OGLE and MACHO surveys exist.

We carried out the analysis of its light curve (from OGLE-III), as well as the period analysis
using all of the available photometry. Our new data from La Silla cover more than 1200~days during
seasons 2012-16. Thanks to all of these data, we were able to reliably identify the period
variation with the period of about seven~years and the surprisingly high amplitude of about
0.040~days, see Table \ref{FigOC}. The resulting mass function of the third body is of about
6.48~M$_\odot$ which, as far as we know, is the highest value among LTTE systems ever studied.
However, even despite its large value, this kind of solution can still be considered a credible
one. It produced the minimal mass of the third body of about 26~M$_\odot$, which is practically the
same as the total mass of the eclipsing pair itself (assumed $M_{12}=26$~M$_\odot$). However, this
is not in contradiction with the light curve solution, because it also resulted in the highest
value of the third light in our sample of stars (see Table \ref{LCparam}) and its value is in
perfect agreement with the predicted third-body mass.

\subsection{OGLE-LMC-ECL-20162}

Another interesting example is the star OGLE-LMC-ECL-20162, which has also been studied only very
briefly in the past. \cite{2012ApJ...748...96M} published its spectral type as O9V. Therefore, we
fixed $T_1=33000$~K for our light curve solution \citep{2013ApJS..208....9P}.

Performing the light curve fitting, as well as the period analysis of the system, we found that
there is also probably the third component with the orbital period of about 5.4~yr. This type of
object (located on the main sequence) should have about the same luminosity as the eclipsing
components. This result is in rough agreement with our light curve solution as presented in Table
\ref{LCparam}.

\begin{figure}
  \centering
  \includegraphics[width=0.45\textwidth]{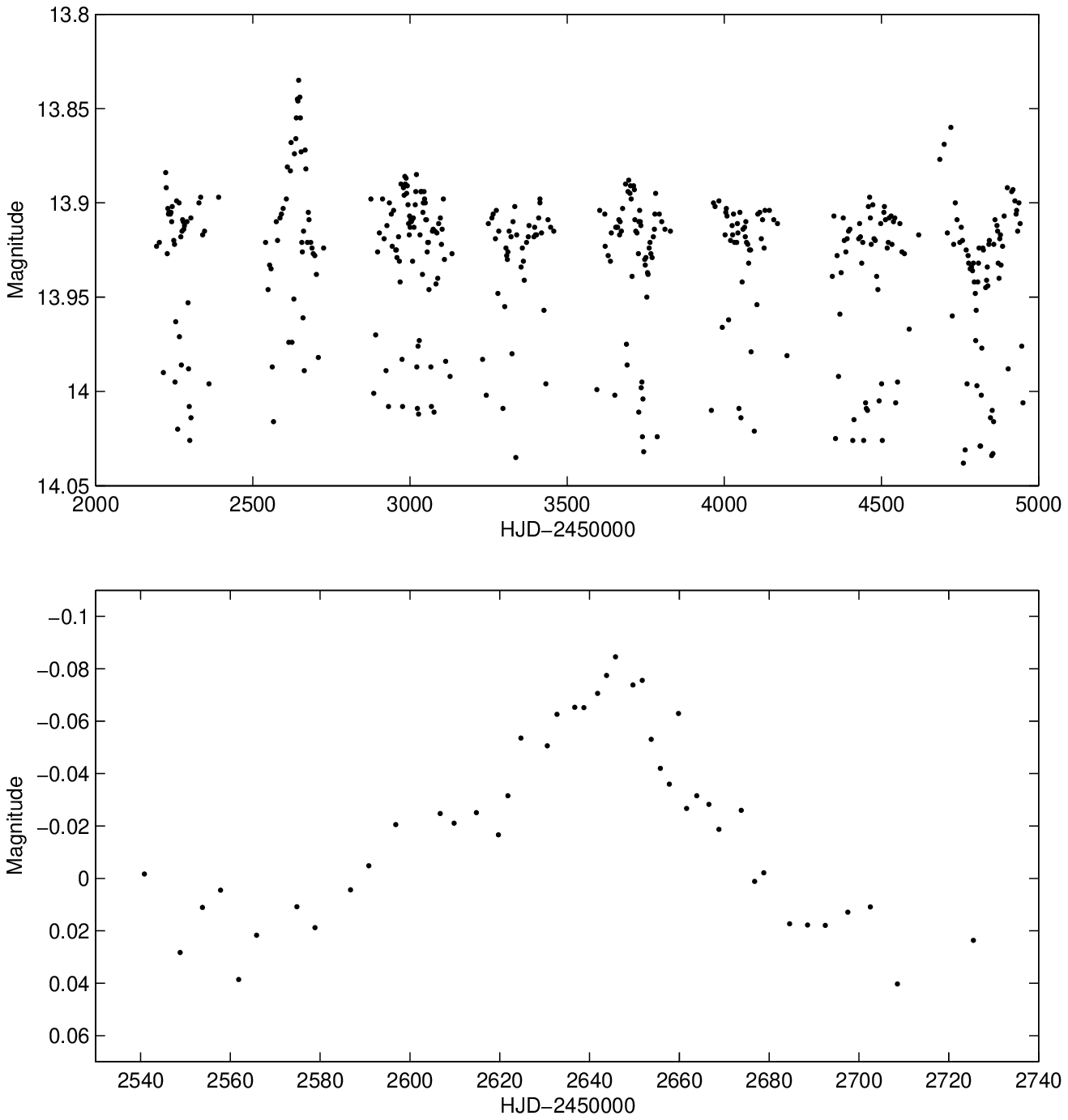}
  \caption{Photometry from the OGLE-III survey of the star OGLE-LMC-ECL-20162 (top). The first
  flare-like/lensing-like event is also plotted and zoomed (bottom figure) but only with the data that ranges from HJD 2452540 to 2452740 after
  subtraction of the light curve fit (i.e. residuals from the solution used for the complete analysis).}
  \label{FigLC20162}
\end{figure}

However, what makes this system more interesting is the fact that some kind of flares were observed
in the light curve. This is most visible in the OGLE data (see Fig. \ref{FigLC20162}), having the
amplitude of about 0.1 mag. However, the MACHO photometry would also indicate some variability of
the light curve shape (however, with no flares). The first event is zoomed in in Fig.
\ref{FigLC20162}, where we also plotted the residuals of the light curve fit. Its nature is still
questionable, because a flare, with such a long duration in such an early-type star is rather
improbable. On the other hand, its shape would also indicate that the gravitational lensing event
is controversial (because it is not perfectly symmetric and its duration is rather long). What
seems to be remarkable is the fact that another flare-like event occurs about 2050 days after the
first one, while the predicted orbital period of the third body is only about 100 days shorter
(forcing the fit to have 2050-day period yielded the $\chi^2$ value to be about 7\% higher).
Therefore, only further analysis would resolve the question about the nature of this phenomenon.

\subsection{OGLE-LMC-ECL-09971}

The last system, which we would like to feature, is the OGLE-LMC-ECL-09971. No analysis can be
found in the literature, hence we deal only with very limited information about the star.

However, the light curve fitting was carried out and the solution is presented in Table
\ref{LCparam}, where we can see that the third light resulted only in a rather small value. But also
the third body, the result of the $O-C$ diagram analysis, has a small mass. On the other
hand what makes this system remarkable is the fact that the individual primary and secondary minima
in the $O-C$ diagram, plotted in Fig. \ref{FigOC}, shows that the orbit could have a non-zero
eccentricity. Residuals after subtracting the LTTE fit show that the secondary minima occur
slightly later than the primaries, however the difference is only very small and only future
dedicated observation or spectroscopy will reveal whether the orbit is slightly eccentric or
circular.

\begin{table*}
\caption{Identification of the analysed systems.}  \label{InfoSystems}
 \footnotesize
\begin{tabular}{lcccccccccccl}
   \hline\hline\noalign{\smallskip}
 System   & OGLE II$^{1}$ &  MACHO     &          RA           &             DE                              & $I_{\rm max}^{2}$ & $(V-I)_0^{3}$ & $(B-V)_0$  \\      
  \hline\noalign{\smallskip}
 OGLE-LMC-ECL-06242 &              & 19.3825.12  & 05$^h$03$^m$46$^s$.71 & -67$^\circ$59$^\prime$27$^{\prime\prime}\!$.5 & 13.859 &  -0.163 & -0.273$^{4}$ \\    
 OGLE-LMC-ECL-08823 &  SC11 331549 &  79.4780.9  & 05$^h$09$^m$40$^s$.43 & -68$^\circ$53$^\prime$24$^{\prime\prime}\!$.8 & 13.543 &   0.035 & -0.219$^{4}$ \\    
 OGLE-LMC-ECL-09971 &  SC9 38527   & 79.5258.1879& 05$^h$12$^m$36$^s$.56 & -69$^\circ$17$^\prime$31$^{\prime\prime}\!$.6 & 16.303 &  -0.173 & -0.241$^{5}$ \\    
 OGLE-LMC-ECL-10140 &  SC9 38636   &  79.5258.87 & 05$^h$13$^m$03$^s$.43 & -69$^\circ$17$^\prime$12$^{\prime\prime}\!$.0 & 16.762 &  -0.224 & -0.145$^{5}$ \\    
 OGLE-LMC-ECL-15313 &  SC4 53802   & 77.7307.358 & 05$^h$25$^m$32$^s$.18 & -69$^\circ$49$^\prime$39$^{\prime\prime}\!$.1 & 16.677 &  -0.113 &  0.063$^{5}$ \\    
 OGLE-LMC-ECL-19996 &              &  81.9003.9  & 05$^h$35$^m$34$^s$.18 & -69$^\circ$39$^\prime$48$^{\prime\prime}\!$.2 & 13.793 &  -0.333 & -0.292$^{5}$ \\    
 OGLE-LMC-ECL-20162 &              &  82.9011.7  & 05$^h$35$^m$55$^s$.06 & -69$^\circ$08$^\prime$55$^{\prime\prime}\!$.0 & 13.906 &  -0.041 & -0.308$^{4}$ \\    
 OGLE-LMC-ECL-20582 &              & 82.9131.137 & 05$^h$36$^m$47$^s$.44 & -69$^\circ$13$^\prime$19$^{\prime\prime}\!$.7 & 16.643 &   0.157 & -0.245$^{4}$ \\    
 \noalign{\smallskip}\hline
\end{tabular}
\\
\begin{minipage}{0.99\textwidth}
 Notes: [1]  The full name from the OGLE II survey should be OGLE LMC-SCn nnnnnn, [2] Value taken from \cite{2011AcA....61..103G}, [3] Value taken from \cite{2012AcA....62..247U}, [4]
Value derived from the (B-V) and (U-B) indices taken from \cite{2002ApJS..141...81M}, [5] Value
taken from \cite{2004AJ....128.1606Z}.
\end{minipage}
\end{table*}

\begin{table*}
\caption{Light curve parameters for the analysed systems, the results from {\sc PHOEBE}.}
\label{LCparam}
 \scriptsize
\begin{tabular}{lccccccccccccc}
  \hline\hline\noalign{\smallskip}
   System & $T_1$ (fixed) & $T_2$  &  $i$ [deg]   & Type & q [$M_2/M_1$] &  $R_1/a$   &  $R_2/a$   & $L_1$ [\%] & $L_2$ [\%] & $L_3$ [\%] \\
  \hline\noalign{\smallskip}
 $\#$06242 &  26000  & 18847 (1096)& 90.84 (1.34) & SD &  3.7 (0.5)    & 0.272 (0.0  ) & 0.294 (0.0  ) &  9.6 (0.7) &  6.1 (0.5) & 84.3 (1.8) \\ 
 $\#$08823 &  79000  & 26876 (1122)& 70.66 (0.26) & D  &  2.67 (fixed) & 0.186 (0.013) & 0.294 (0.016) & 39.5 (1.0) & 22.8 (0.6) & 37.7 (2.7) \\ 
 $\#$09971 &  24000  & 21143 (117) & 79.95 (0.16) & D  &  0.99 (0.05)  & 0.290 (0.009) & 0.349 (0.011) & 46.1 (1.2) & 52.2 (0.9) &  1.7 (0.9) \\ 
 $\#$10140 &  15000  & 14954 (239) & 79.00 (0.25) & D  &  0.96 (0.02)  & 0.190 (0.004) & 0.176 (0.004) & 52.3 (0.7) & 44.5 (0.6) &  3.1 (1.1) \\ 
 $\#$15313 &  13000  &  7376  (80) & 85.54 (0.73) & SD & 13.3 (0.7)    & 0.188 (0.010) & 0.301 (0.009) & 25.9 (1.9) & 18.5 (2.7) & 55.6 (4.0) \\ 
 $\#$19996 &  28000  & 26099 (609) & 68.85 (0.56) & D  & 0.88 (0.13)   & 0.307 (0.008) & 0.281 (0.011) & 27.6 (1.7) & 20.6 (1.5) & 51.8 (2.4) \\ 
 $\#$20162 &  33000  & 27295 (746) & 68.18 (0.58) & D  & 0.74 (0.03)   & 0.322 (0.007) & 0.243 (0.006) & 50.4 (1.2) & 23.1 (2.3) & 26.5 (1.7) \\ 
 $\#$20582 &  24500  & 16706 (248) & 76.00 (0.19) & D  & 0.55 (0.05)   & 0.326 (0.008) & 0.247 (0.005) & 66.7 (0.8) & 21.3 (0.8) & 12.0 (1.1) \\ 
 \noalign{\smallskip}\hline
\end{tabular}
\end{table*}

\begin{table*}
\caption{The parameters of the third-body orbits for the individual systems.} \label{OCparam}
 \scriptsize
\begin{tabular}{ccccccccccc}
\hline\hline\noalign{\smallskip}
   System        & $HJD_0     $  &   $P$      &  $A$         & $\omega$   & $P_3$      & $T_0$ [HJD]   &      e      & $f(m_3)$   & $P_3^2/P$ \\   
                 & (2450000+)    &   [days]   & [days]       &  [deg]     & [yr]       & (2400000+)    &             & $[M_\odot]$& [yr]      \\
 \noalign{\smallskip}\hline\noalign{\smallskip}
 $\#$06242 & 52000.6153 (74) & 2.0154961 (50) & 0.0201 (54) & --          & 12.7 (1.7) & 2449306 (407) & 0.0         & 0.262 (15) & 29238 \\ 
 $\#$08823 & 52006.6737 (260)&17.9947441 (200)& 0.0831 (255)&188.2 (31.0) &  8.9 (1.4) & 2461802 (675) & 0.34 (0.21) & 45 (16)    & 1611  \\ 
 $\#$09971 & 53560.8965 (39) & 2.3012716 (19) & 0.0082 (35) &203.2  (8.9) &  3.6 (0.3) & 2451990 (145) & 0.29 (0.07) & 0.245 (31) & 2057  \\ 
 $\#$10140 & 53566.8102 (112)& 3.2893145 (114)& 0.0076 (16) & 10.5  (8.5) &  8.7 (0.4) & 2454371  (98) & 0.64 (0.27) & 0.063 (37) & 8437  \\ 
 $\#$15313 & 53562.4212 (27) & 2.5341031 (26) & 0.0182 (40) & --          &  8.6 (0.4) & 2457853 (118) & 0.0         & 0.425 (22) &10590  \\ 
 $\#$19996 & 56975.5993 (54) & 1.0794168 (15) & 0.0402 (23) & --          &  7.2 (0.2) & 2453300  (49) & 0.0         & 6.481 (109)&17456  \\ 
 $\#$20162 & 52002.5561 (33) & 3.9125347 (71) & 0.0227 (33) & --          &  5.4 (0.2) & 2453009  (58) & 0.0         & 2.077 (132)& 2699  \\ 
 $\#$20582 & 53571.3945 (14) & 1.7765410 (10) & 0.0102 (11) & --          &  6.6 (0.1) & 2452925  (38) & 0.0         & 0.127 (10) & 8827  \\ 
 \noalign{\smallskip}\hline
\end{tabular}
\end{table*}

\begin{table}
 \tiny
 \scalebox{0.9}{
 \begin{minipage}{90mm}
  \caption{Heliocentric minima of the systems used for the analysis.} \label{MINIMA}
   \tiny
  \begin{tabular}{@{}l l l l l l@{}}
\hline
Star & HJD - 2400000 & Error & Type  & Filter & Source\\
 \hline
 $\#$06242 & 52260.60572 &  0.01422 &  Prim &  I &  OGLE III \\
 $\#$06242 & 52261.62317 &  0.02129 &  Sec  &  I &  OGLE III \\
 $\#$06242 & 52558.89680 &  0.00358 &  Prim &  I &  OGLE III \\
 $\#$06242 & 52970.07125 &  0.00332 &  Prim &  I &  OGLE III \\
 $\#$06242 & 52971.08682 &  0.00648 &  Sec  &  I &  OGLE III \\
 $\#$06242 & 53283.49384 &  0.00826 &  Sec  &  I &  OGLE III \\
 $\#$06242 & 53696.67741 &  0.00595 &  Sec  &  I &  OGLE III \\
 $\#$06242 & 54153.18234 &  0.00453 &  Prim &  I &  OGLE III \\
 $\#$06242 & 54154.20556 &  0.02131 &  Sec  &  I &  OGLE III \\
 $\#$06242 & 54725.54243 &  0.00598 &  Prim &  I &  OGLE III \\
 $\#$06242 & 54726.58087 &  0.00645 &  Sec  &  I &  OGLE III \\
 $\#$06242 & 49027.76347 &  0.01304 &  Prim & BR &  MACHO \\
 $\#$06242 & 49028.80096 &  0.02138 &  Sec  & BR &  MACHO \\
 $\#$06242 & 49376.45189 &  0.01611 &  Prim & BR &  MACHO \\
 $\#$06242 & 49377.51490 &  0.04892 &  Sec  & BR &  MACHO \\
 $\#$06242 & 49624.36884 &  0.00614 &  Prim & BR &  MACHO \\
 $\#$06242 & 49625.36958 &  0.03671 &  Sec  & BR &  MACHO \\
 $\#$06242 & 49874.26455 &  0.00683 &  Prim & BR &  MACHO \\
 $\#$06242 & 49875.29005 &  0.07545 &  Sec  & BR &  MACHO \\
 $\#$06242 & 50126.23542 &  0.01035 &  Prim & BR &  MACHO \\
 $\#$06242 & 50127.20754 &  0.01421 &  Sec  & BR &  MACHO \\
 $\#$06242 & 50376.12807 &  0.02559 &  Prim & BR &  MACHO \\
 $\#$06242 & 50377.12709 &  0.02189 &  Sec  & BR &  MACHO \\
 $\#$06242 & 50696.58725 &  0.02066 &  Prim & BR &  MACHO \\
 $\#$06242 & 50697.62444 &  0.01760 &  Sec  & BR &  MACHO \\
 $\#$06242 & 51226.63597 &  0.03305 &  Prim & BR &  MACHO \\
 $\#$06242 & 51227.64686 &  0.04246 &  Sec  & BR &  MACHO \\  \hline
 \dots \\
  \hline
 \end{tabular}
 This table is available in its entirety as a machine-readable table. A portion is shown here for
 guidance regarding its form and content.
\end{minipage}}
\end{table}

\section{Results and discussion }

The methods for analysing eclipsing binaries are nowadays classical and used almost routinely.
However, this kind of analysis can still bring new and surprising results, especially when we apply
these methods to new kind of binaries. This is the case of this study, which presents an analysis
of the period changes of binaries outside of the Milky Way Galaxy.

We applied classical LTTE hypothesis to eight selected eclipsing systems of early spectral type
from the LMC and found that third components probably exist with rather short orbital periods of a
couple of years. The significant selection effect has a role because we only deal with limited
amounts of data points that cover about 20~years (and, moreover, the selection also has an effect
on the sample of binaries, i.e. only massive ones). This is the first study of its kind of
extragalactic sources and we have shown that this type of analysis is also suitable for the stars
outside of our Galaxy.

People may have quesitons about the quality of the input data and the conclusiveness of the LTTE fit. We are
aware that some of the minima times have rather poor quality and their respective
uncertainties are comparable with the amplitudes of the LTTE (this applies mostly for the old MACHO
data). One the other hand, it is only because of these older data points that we are able to identify the
period changes with longer periods.

The analysis has shown that the predicted third bodies found via the period changes have rather
high masses in general, but this is also due to the fact that the eclipsing binary components are
very massive ones. We deal here with the stars of the highest luminosity, the highest mass, and of
the earliest spectral type in the LMC and, consequently, the third bodies should also be massive.
This finding was supported by the fact that large fractions of the third light were also detected
in the light curve solution and these two numbers coincide well with each other for most of the
systems.

Another finding from our study is that the dynamical effects of the third bodies on the eclipsing
binary orbits are generally small. This can be seen in the last column in Table \ref{OCparam} where
the ratio of periods can tell us something about the influence of the third component on the inner
orbit and the ratio $P_3^2/P$ can be considered as a typical timescale of the precession caused by
the distant body. As a result, we cannot hope to find any evidence there of precession on a
timescale of decades, but rather centuries. This is caused by the inner binary having a relatively
short orbital period while, on the contrary, the third body has too long orbital period.

And finally, another remarkable finding is the fact that in our study of eight systems,
 five orbits were discovered to be circular. If we state that the solution for the system
OGLE-LMC-ECL-08823 is odd and cannot be considered as a correct one, hence we deal with of about
two-thirds of all circular systems. This result is quite surprising, because our experience of
analysing many dozens of LTTE orbits tells us that these systems are mostly eccentric. Also other
studies of LTTE in another sets of stars (e.g. \citealt{2015AJ....149..197Z}, or
\citealt{2016MNRAS.455.4136B}) tend to show eccentric solutions. Whether this is also the result of
some selection effect or some mechanism of faster circularization for the very massive stars is
still not clear.

\section{Conclusion}

A set of eight luminous eclipsing binaries in LMC were analysed resulting in the finding that all
these systems probably contain some additional components. Despite the fact that this is still only
a hypothesis, we can consider this finding as a good starting point for future dedicated
observations of these systems. Especially those systems like OGLE-LMC-ECL-19996 or
OGLE-LMC-ECL-20162 seem to be the most appropriate for follow-up spectroscopic observations because
these are relatively bright enough stars, their predicted third components are luminous enough when
compared with the eclipsing pairs, and finally the periods of the outer orbits are also relatively
short.

\begin{acknowledgements}
An anonymous referee is acknowledged for helpful and critical suggestions which significantly
improved the manuscript. We thank the {\sc MACHO} and {\sc OGLE} teams for making all of the public
observations easily available. This work was supported by the Czech Science Foundation grants no.
P209/10/0715, and GA15-02112S, and also by the grant MSMT INGO II LG15010. We are also grateful to
the ESO team at the La Silla Observatory for their help in maintaining and operating the Danish
telescope. The following internet-based resources were used in research for this paper: the SIMBAD
database and the VizieR service operated at the CDS, Strasbourg, France, and the NASA's
Astrophysics Data System Bibliographic Services.
\end{acknowledgements}

\end{document}